# Infant Vocal Tract Development Analysis and Diagnosis by Cry Signals with CNN Age Classification

*Chunyan Ji[1], Yi Pan[2]*
[1]Department of Computer Science, Georgia State University, Atlanta, USA
[2]Shenzhen Institute of Advanced Technology, Chinese Academy of Sciences, Shenzhen, China

*Abstract*—From crying to babbling and then to speech, infants' vocal tract goes through anatomic restructuring. In this paper, we propose a non-invasive fast method of using infant cry signals with convolutional neural network (CNN) based age classification to diagnose the abnormality of the vocal tract development as early as 4-month age. We study F0, F1, F2, and spectrograms and relate them to the postnatal development of infant vocalization. A novel CNN based age classification is performed with binary age pairs to discover the pattern and tendency of the vocal tract changes. The effectiveness of this approach is evaluated on Baby2020 with healthy infant cries and Baby Chillanto database with pathological infant cries. The results show that our approach yields 79.20% accuracy for healthy cries, 84.80% for asphyxiated cries, and 91.20% for deaf cries. Our method first reveals that infants' vocal tract develops to a certain level at 4-month age and infants can start controlling the vocal folds to produce discontinuous cry sounds leading to babbling. Early diagnosis of growth abnormality of the vocal tract can help parents keep vigilant and adopt medical treatment or training therapy for their infants as early as possible.

## I. Introduction

Novice parents are excited to hear their newborn's first cry and care about their health the most. Infants express their needs, such as pain, discomfort, and hunger, etc., by crying. It is shown that the postnatal development of vocal tract is associated with cry signals [1]. Diseases can lead to vocal tract development retardation and some healthy infants may also suffer from vocal tract development delay. Many studies explore the anatomical and acoustic features of adult vocal tract, only a few for children, and even less studies are for infants. Medical methods such as computed tomography (CT) and magnetic resonance imaging (MRI) techniques discover anatomical vocal tract development. It is known that an infant's vocal tract is not simply a miniature version of an adult's vocal tract [1]. Previous research has shown that infant vocal tract increases more than twofold in length and its geometric proportions also change [2]. The shape of the infant vocal tract changes from infancy to adulthood. In the process of the vocal tract development, the bend in the oropharyngeal region gradually forms a right angle, both larynx and the posterior part of the tongue descend and the distance between the soft palate and epiglottis is enlarged. Fitch and Giedd study MRI images of subjects from 2 to 25 years old and point out the first phase larynx descend occurs early in life and the second large descend, which is restricted to males, occurs at puberty [3]. Researchers discovered that the acoustic features of infant vocalization reflect the changes in the vocal tract. Kent and Murray discovered the ranges of both F1 and F2 frequencies increase as infant grow from 3-month to 6-month age [4]. Machine learning methods have been used in studying infant growth. Pruett et al. performed age classification on 6 versus 12-month old infants by functional connectivity magnetic resonance imaging (fcMRI) data to study brain and behavioral development using support vector machine (SVM) [5].

Speech emergence and development is presumed to be dependent, at least in part, on the physical changes that the vocal tract structures undergo during development [2]. Speech development starts as early as infant crying. Robb et al. confirmed the production of laryngeal constriction during infants' 3-5 months supports the notion that infants start to test and practice their phonetic production skills in the first several months of life [6]. Guiding infants, especially infants with tardy vocal tract development, to practice certain sounds and syllables as early as possible promotes their speech development. Finding out whether infants' vocal tract is developing normally as expected is vital for parents to take timely measures against the problems found. Compared to MRI with image processing, infant cry analysis and classification is non-invasive. The combination of signal processing and machine learning technologies on portable devices leads to simple and easy procedures, which can be performed without professionals.

In this study, we analyze diverse cry signals from different monthly age of infants and extract typical features such as F0, F1, F2, and spectrograms to investigate the relationship between anatomic changes of vocal tract and the characteristics of cries. Through this study, we discover that 4-month age is a key turning point of infant vocal tract growth. Moreover, we apply efficient neural networks to discover the pattern of the changes and diagnose the abnormality of the infant vocal tract by age classification. To our best knowledge, this is the first work of age classification via classifying infant cry signals. In this paper, our major contributions include:

- We propose using the characteristics of infant cry signals to evaluate the development of vocal tract development.

- Fundamental frequency (F0), formants (F1 and F2), and spectrograms of infant cries are investigated and related to the postnatal development of infant vocalization and we show that 4-month age is a key turning point of infant vocal tract development.

- An efficient convolutional neural network (CNN) approach is applied to infant cry binary monthly age classification to discover the trend of infants' vocal tract development and diagnose abnormal vocal tract development as early as 4-month-old.

The remainder of the paper is organized as follows. In Section 2, acoustic features of infant cry signals are analyzed to confirm the relationship between infant cries and infant vocal tract development. Section 3 describes the CNN architecture used for age classification in this study. In Section 4, experimental results of monthly age classification and diagnosis of the abnormality of vocal tract development are presented, and we conclude in Section 5.

## II. INFANT VOCAL TRACT ANALYSIS

### A. General Description of Infant Vocalizations

The shape of the newborn's vocal tract is more like a chimpanzee than a human adult [7]. The high position of the larynx and Hyoid bone causes the difficulty of controlling the tongue making infants unconducive to pronunciation. In the postnatal development phase, infants' vocal tract restructures gradually and develops mature speech ability accordingly. From infancy to adulthood, the length of the vocal tract develops from about 8cm to 17cm. It is shown that the vocal tract is nonlinear gradual and the growth curve of it can be fitted with fourth degree polynomial model [1]. Figure 1 shows the changes of the vocal tract from infants to adults. In the process of the vocal tract development, the bend in the oropharyngeal region gradually forms a right angle. Both infants' larynx and the posterior part of the tongue descend and the distance between the soft palate and epiglottis is enlarged. Hence, the infant vocalization is without resonant effect and the vowels sound within cries is nasalized resulting in quite different distribution of F1 and F2.

Infants' speech development starts from crying. Previous research shows that healthy infants cry for around 1.75 hours per day by the second week of life, reaches the peak of 2.75 hours by 6 weeks, and decreases gradually to 0.75 hours by 12 weeks [8]. Figure 2 presents a comparison of time and frequency domain between infant cry and adult speech. The areas included in the green rectangles are basic cries and the ones in the blue rectangles are cries ending with creaks. The spectrum of basic cries is with clear bars, which is similar to that of vowels (green rectangles) shown in the Figure 1b. The creak at the end of a cry is like choking or interruptions with no vibration of vocal cord. Comparing to infants' cry signals, adults' speech signals are more complex and richer with energy, intensity, and formants changes representing a variety of the expressions. Because of infants' lack of full control of the vocal tract, they can only control the breath force from the lung to generate different types of cries for diverse purposes. The effect of movement of vocal cords is based on Bernoulli's effect [9]. Bernoulli effect determines the movement of the vocal cords to present such characteristics as the higher the flow rate, the lower the pressure. The flow rate increases when air comes from the lungs and passes through the narrow glottis. According to the Bernoulli's principle, the pressure at the vocal cords is reduced and the vocal cords are closed, and then the subsequent air opens the vocal cords again. Consequently, the sound is produced because the vocal cords keep moving up and down repeatedly. The harder the infant breathes, the faster the frequency of the opening and closing of the vocal cords and the greater the pitch and loudness of the sound.

### B. Analysis of Different Monthly Age Infant Cries

A spectrogram is a visual representation of an audio signals showing the amplitude of a particular frequency at a particular time. The spectrograms shown in Figure 3 illustrate

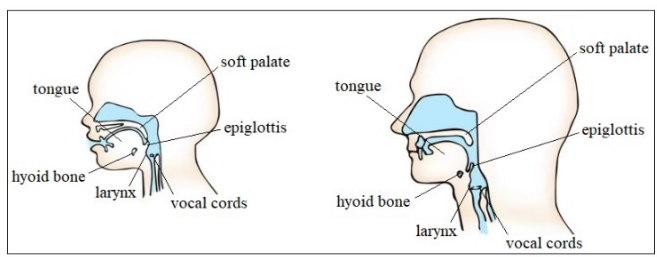

Fig. 1. comparison of vocal tract structure of newborn and adult.

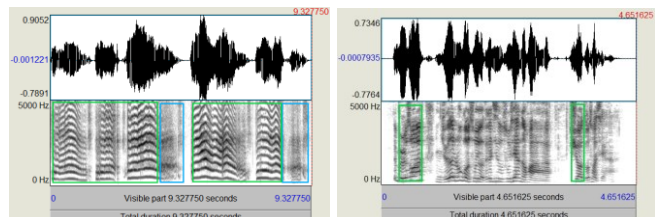

(a): Waveform and spectrogram of an infant cry

(b): Waveform and spectrogram of an adult speech the difference between a common 1-month cry and 4-month cry. In the earliest three months, an infant cry is characterized by its periodic nature, which alternates crying and inspirations. We can see that the clear harmonics are both in the lower frequency region below 3KHz, which covers more energy represented by lighter colors in the spectrograms. The harmonic structure becomes drastically weaker as the frequency increases for both spectrograms, but the 4-month spectrogram contains stronger energy in the low frequency than the 1-month spectrogram. Figure 3(b) illustrates a gap, around 0.85 second, which is the effect of glottis closure. When we listen to this audio signal, it contains an unclear "mama" sound. It means the 4-month age infant acquires the ability to close the glottis to form a certain level discontinuous vocalization during crying. It means the infant can generate discontinuous speech with different blocks in a whole articulation and is ready for the first word pronunciation.

Fig. 2. Waveform and spectrogram of infant cry and an adult speech.

The shape of vocal tract decides the resonant characteristics of its vocalizations. Infant cry signal is characterized by its high F0 within 250-700Hz compared to 85Hz to 200Hz of adult. The first two formants (F1 and F2) determine the vowel sounds, relating to the length and place of narrowing of the vocal tract and the F0 is corresponding to the increases in the length and volume of a vocal cord [1]. F1 corresponds to the vertical height (high or low) of the tongue, F2 relates to the horizontal position (forward and backward) of the tongue [1]. In Figure 4, we plot F0, F1, and F2 using Praat [10] tool for a typical male infant cry for hunger at 1-month age and 4-month age. It is shown that the coefficients of F0 relating to vocal cord vary slightly between 1-month age and 4-month age baby. It indicates that the length and the volume of the vocal cord may not change much during the postnatal development of the first 3 months. On the other hand, values of F1 and F2 (F1=1921Hz vs 1470Hz, F2=4423Hz vs 2339 Hz) for 4-month vocalization increased significantly, which is in accordance with our previous analysis. Since F1 and F2 are strongly related to resonant cavity and the tongue, the increase of F1 and F2 indicates a great change of tongue location, oral, and nasal cavity extension for word pronunciation.

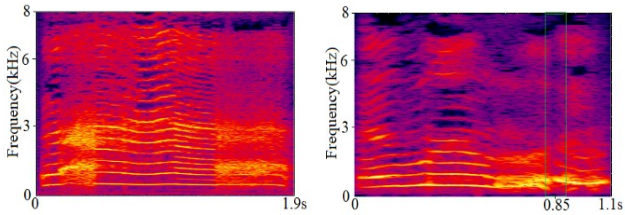
(a): 1-month sleepy cry  (b): 4-month sleepy cry with "mama"
Fig. 3. Spectrograms of a 1-month sleepy cry and a 4-month sleepy cry.

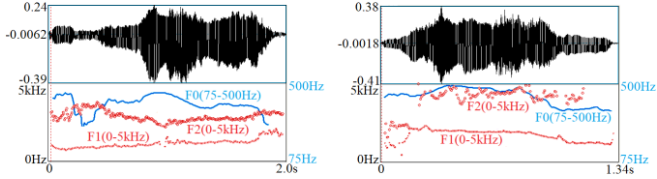
(a): 1-month infant hungry cry  (b): 4-month infant hungry cry
F0=442Hz, F1=1470Hz, F2=2339Hz  F0=457Hz, F1=1921Hz, F2=4423Hz
Fig. 4. F0, F1, F2 comparisons of same infant at 1-month and 4-month.

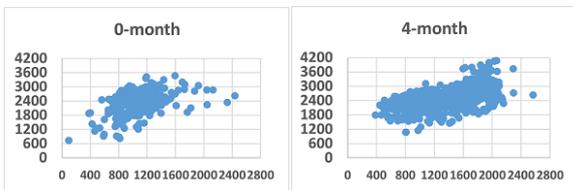
Fig. 5. F1 and F2 distribution from a certain infant of 0-month and 4-month.

To show the distribution of F1 and F2 along with the vocal tract development, we plot the scatter graphs for 0-month and 4-month cry samples from the same boy infant in Figure 5. Each graph contains the same number of values extracted from 100 cry samples. The horizontal axis is the F1 value, and the vertical axis is the F2 value. Figure 5 indicates that with the development of vocal tract, the average values of F1 and F2 increase. For example, F1 over 800Hz and F2 over 2400Hz are covered by more samples from 4-month compared to those of 0-month. In addition, the distribution of samples from 4-month is quite different with respect to 0- month samples. The standard deviation of F1 and F2 are increased with age. It indicates that with the improved ability of controlling the vocal tract, infants start to generate different formants changes representing different expressions at 4-month age, which is a turning point of the development of vocal tract.

## III. AGE CLASSIFICATION WITH CONVOLUTIONAL NEURAL NETWORKS

Convolutional neural network (CNN) is one of the deep learning models that is widely used in many research domains such as image classification, object detection, and signal processing, etc. Comparing to fully connected neural network, CNN is better at extracting the features from the images with less parameters to train, which leads to better performance and less training time. In the training phase, each labeled image passes through a certain number of convolutional layers with selected filters, selected pooling that reduces the dimensionality, fully connected layers, and at last a softmax activation function that is applied to the last dense layer to generate a probabilistic value between 0 and 1 for classification. While training, the filter weights get updated by backpropagation algorithm to ensure the result matches the label of the image. In the testing phase, testing images pass through the trained CNN model to get the classification labels.

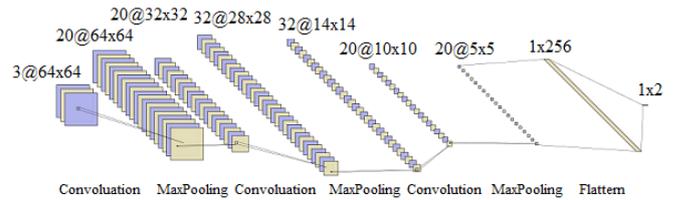
Fig. 6. CNN architecture of our approach for age classification.

In recent years, CNN is used in infant cry reason classification and infant cry detection. The input images used include waveforms, spectrograms, and prosodic feature images. Research shows that the spectrograms perform the best on classifying the cry signals comparing to waveforms and prosodic line images [11]. In this study, we extract the spectrograms from the cry signals and feed them into the CNN model for age classification. The implementation of the CNN uses Keras framework with Tensorflow backend [12]. The architecture of our CNN model is illustrated in Figure 6. Spectrograms are fed into the CNN model, which contains three convolutional layers, maxpooling, and the network is flattened into a 256-neuron fully connected layer, and then the softmax is used in the last layer for classification.

## IV. EXPERIMENTAL SETUP AND RESULTS

### A. Dataset

The dataset used in this study is a subset of the developing Baby2020 database. Baby2020 database samples are collected from over 100 babies from newborn to 9 months old via mobile devices placed right beside the infants in natural real-world home or hospital environments. The length of each recording is less than 3 minutes. Cry samples are manually segmented with the length between 1 second to 7 seconds using the Transcriber tool [13]. Cry samples are labeled either by parents at home or by doctors and nurses in hospitals. Each recording is labeled by the monthly age of the infant, gender of the infant, and reason of the cry. The cry reason annotation is based on caregivers' experiences and the real time situation when the cry occurs.

In this study, we use cry samples from infants between newborn (0-month) to 6-month. There are 10500 samples in total, including 1500 samples from multiple infants and multiple types of the cries selected for each month. The types of the cry samples include hungry, sleepy, unhappy, wakeup, attention, uncomfortable, temper, and pain. The healthy cry testing set for the abnormal vocal tract development diagnosis experiment contains 220 samples for each month. The deaf and asphyxiated testing set is from the Baby Chillanto database [14], which contains the cries from babies ranging from newborn to 9-month of age and each sample is a one-second long audio wav file. We use the 340 deaf cry samples and 879 asphyxiated cry samples as the abnormal cry samples for our experiments. We use Sound eXchange (Sox) software [15] to generate the spectrograms and images being fed into the CNN are resized into 64*64.

### B. Experimental setup and results

A CNN binary classifier is used to discover the pattern of the changes of the vocal tract. There are three convolutional layers followed by three max-pooling layers in the model. 5*5 filter size, 2*2 pooling size and 2*2 stride are used. The first convolutional layer uses "same" padding, and the default "valid" padding is used in other two convolutional layers. The first and third convolutional layers use 20 filters, and the second convolutional layer uses 32 filters. After the third max-pooling, the network is flattened into a 256-neuron dense layer. The optimizer used is Adam, and the ReLu activation function is used in each convolution layer. In the dense layer, the softmax activation function is applied for final classification. 5-fold cross validation is performed on the identifying vocal tract development experiments as training and testing samples are from the same group of infants. The testing samples and training samples for diagnosing abnormal vocal tract are from different group of infants. All models are trained with 100 epochs and all testing accuracies are the average of 10 runs.

*1) Identify infant vocal tract development by binary age classification.* In Section 2 we analyze that 4-month age is a vital turning point of vocal tract development. We perform binary age classification on infant cry signals to experiment this change. The labels of the samples are 0, 1, 2, 3, 4, 5 and 6 representing the monthly age of the infants. We generate different binary pairs such as 01, 02, 03, 04, 05, 06, 12, 13, 14, 15, 16, 23, 24, 25, and 26. For example, 01 means classifying 0-month samples from 1-month samples. Table I and Figure 7 give the classification results of all pairs. CNN achieves over 85% for all pairs indicating its strong ability to differentiate the monthly cries. As shown in Figure 7, when 0-month, 1-month, and 2-month cries are compared to cries in other months, the accuracies consistently increase as infants grow. The turning points arrive at 4-month where the accuracies stop increasing or frustrates through 6-month. The classifiers cannot differentiate the cries of 5-month and 6-month cries from 0-2-month cries better than the 4-month cries indicating the change of the infant vocal tract reaches a certain stable stage after the infants reach 4-month old. In another experiment, we separate the male cries and female cries into two datasets and perform the binary age classification separately, we discover that both datasets show the same trend as the combined dataset shows, which is the accuracies stops increasing when infants reach 4-month-old. But all classification accuracies of the male cries are lower than the ones for female cries. It may indicate that the vocal tract development of the male infants is slower than the female infants from newborn to 5 months.

*2) Abnormal vocal tract development diagnosis.* We perform age classification to diagnose the abnormality of the infant vocal tract development. When an infant reaches 4-month and his cry signals are classified as younger month cries, it indicates that his vocal tract development is in growth retardation or related diseases may be involved. We use the same CNN binary classification model described above. We consider month 0, 1, 2, 3 cries as one category and month 4 as another category with 1500 samples in each category. Table II shows the accuracies of the abnormality diagnosis on healthy infants, asphyxiated infants, and deaf infants. For healthy infant cries, 79.20% testing samples can be classified correctly. Asphyxiated infant cries and deaf infant cries are both abnormal cries, especially the deaf infants. The result shows that 84.80% asphyxiated infant cries are classified as cries younger than 4-month cries, and 91.20% of the deaf cries are diagnosed as cries younger than 4-month cries. The loss of hearing of deaf infants has great impact of the Monthly binary age classification accuracies development of vocal tract. It is essential to perform treatment as early as possible. Meanwhile, asphyxiated infants with the pathological effects show the delay of development of vocal tract. The experimental results of these two different types of abnormal pathological cries indicate that our proposed diagnosing classifier is effective.

TABLE I. MONTHLY BINARY AGE CLASSIFICATION ACCURACIES

|  | 0-month | 1-month | 2-month |
|---|---|---|---|
| 1-month | 88.69% | -- | -- |
| 2-month | 93.47% | 88.16% | -- |
| 3-month | 95.91% | 89.98% | 85.16% |
| 4-month | 96.27% | 92.33% | 90.14% |
| 5-month | 96.07% | 93.68% | 90.01% |
| 6-month | 95.70% | 92.58% | 90.22% |

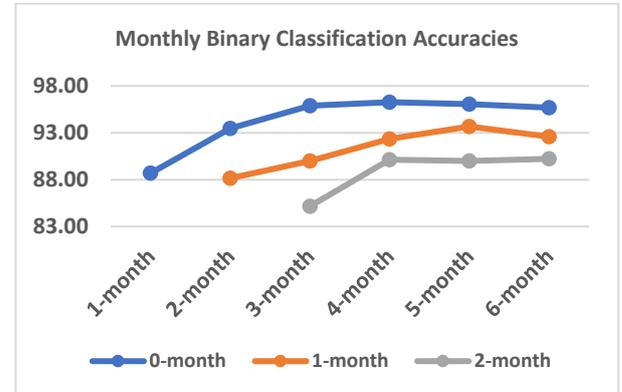

Fig. 7. Line chart of monthly binary age classification.

TABLE II. ACCURACIES OF ABNORMAL VOCAL TRACT DEVELOPMENT DIAGNOSIS BY INFANT CRIES

| CNN Model | Testing samples | Accuracy |
|---|---|---|
| 4 month or younger | Healthy cries (1100) | 79.20% |
| 4 month or younger | Asphyxiated cries (879) | 84.80% |
| 4 month or younger | Deaf infants' cries (340) | 91.21% |

V. CONCLUSION AND FUTURE WORK

In this paper, we demonstrated that infant cry age classification with CNN is an efficient non-invasive method to diagnose the abnormality of the vocal tract development of early age infants. The analysis of acoustic features from different monthly age infants shows that 4-month age is a key turning point of infant vocal tract growth. We have shown that the length and volume of vocal cord may not change much during the postnatal development of the first 3 months. F1 and F2 increase significantly within the early 4 months indicating a great change of tongue location, oral, and nasal cavity extension for word pronunciation. We applied CNN to discover the pattern of the changes and diagnose the abnormality of the infant vocal tract by age classification. Our method achieved 79.20% accuracy for healthy infant cries on Baby2020 database, obtained 84.80% of asphyxiated cries, and 91.20% of the deaf cries for abnormal vocal tract development diagnosis on Baby Chillanto database. In the future, we will apply other machine learning methods, such as SVM and graph neural networks, to improve the age classification accuracy. We plan to expand the dataset to include older infant cries and young children's early speech to study the vocal tract development in children's first several years of growth. With large dataset and the combination of different audio features, we plan to apply rough set theory [16][17] into our audio classification for better performance in time and accuracy.

ACKNOWLEDGMENT

The authors acknowledge molecular basis of disease (MBD) at Georgia State University for supporting this research and the support of NVIDIA Corporation with the donation of the Tesla K40 GPU for providing GPU resource. We like to thank Dr. Carlos A. Reyes-Garcia, Dr. Emilio Arch-Tirado and his INR-Mexico group, and Dr. Edgar M. Garcia-Tamayo for their dedication of the collection of the Infant Cry database. We also want to express our great gratitude to Dr. Orion Reyes and Dr. Carlos A. Reyes for providing the access to the Baby Chillanto database. We thank all parents, doctors, and nurses who support the recording of Baby2020 database.